\documentclass[aps,prd,twocolumn,superscriptaddress,showpacs]{revtex4}
\usepackage{graphicx}
\usepackage{amsmath}
\usepackage{amsfonts}
\usepackage{amssymb}

\begin{document}
\title{Exploring Lee-Wick finite electrodynamics}
\author{Antonio Accioly}\email{accioly@cbpf.br}
\affiliation{Laborat\'{o}rio de F\'{\i}sica Experimental (LAFEX), Centro Brasileiro de Pesquisas F\'{i}sicas (CBPF), Rua Dr. Xavier Sigaud 150, Urca, 22290-180, Rio de Janeiro, RJ, Brazil}
\affiliation{Instituto de F\'{\i}sica Te\'{o}rica (IFT), S\~{a}o Paulo State University (UNESP), Rua Dr. Bento Teobaldo Ferraz 271, Bl. II-Barra Funda, 01140-070 S\~{a}o Paulo, SP, Brazil}
 \author{Patricio Gaete}\email{patricio.gaete@usm.cl}
 \affiliation{Departmento de F\'{i}sica and Centro Cient\'{i}fico-Tecnol\'ogico de Valpara\'{i}so, Universidad T\'{e}cnica Federico Santa Mar\'{i}a, Valpara\'{i}so, Chile}
\author{Jos\'{e} Helay\"{e}l-Neto}\email{helayel@cbpf.br}
\affiliation{Laborat\'{o}rio de F\'{\i}sica Experimental (LAFEX), Centro Brasileiro de Pesquisas F\'{i}sicas (CBPF), Rua Dr. Xavier Sigaud 150, Urca, 22290-180, Rio de Janeiro, RJ, Brazil}
\author{Eslley Scatena}\email{scatena@ift.unesp.br}
\affiliation{Instituto de F\'{\i}sica Te\'{o}rica (IFT), S\~{a}o Paulo State University (UNESP), Rua Dr. Bento Teobaldo Ferraz 271, Bl. II-Barra Funda, 01140-070 S\~{a}o Paulo, SP, Brazil}
\author{Rodrigo Turcati}\email{turcati@cbpf.br}
\affiliation{Laborat\'{o}rio de F\'{\i}sica Experimental (LAFEX), Centro Brasileiro de Pesquisas F\'{i}sicas (CBPF), Rua Dr. Xavier Sigaud 150, Urca, 22290-180, Rio de Janeiro, RJ, Brazil}


\begin{abstract}
We consider the Lee-Wick (LW) finite electrodynamics, i.e., the $U(1)$ gauge theory where a (gauge-invariant) dimension-6 operator containing higher-derivatives is added to the free Lagrangian of the $U(1)$ sector. Three bounds on the LW heavy photon mass are then estimated. It is amazing that one of these bounds, actually  the most reliable one, is of the order of the vectorial bosons masses found in nature. The lowest order modification of the Coulomb potential due to the presence of the higher-derivative term is obtained afterward by means of two outstanding methods: one of them is based on the marriage of quantum mechanics with the nonrelativistic limit of quantum field theory; the other, pioneered by Dirac,  makes use of   a gauge-invariant but path-dependent variables formalism. Interestingly enough, these approaches, despite being radically different, lead to the same result which seems to indicate that they are equivalent term by term.       

\end{abstract}\pacs{14.70.-e, 12.60.Cn, 13.40.Gp}
\maketitle
\section{Introduction}
In the early 1970s, Lee and Wick proposed a finite theory of QED \cite{1,2}. Just about two years ago, Grinstein, O'Connell and Wise, building on pioneering work of these authors, introduced non-Abelian LW gauge theories \cite{3}. Their model, usually referred as the LW Standard Model (LWSM), is naturally free of quadratic divergences, thus providing an alternative way to the solution of the hierarchy problem. Nevertheless, unlike the original LW model, the LWSM  is not a finite theory; yet it is still renormalizable. We call attention to the fact that higher-dimensional operators --- containing new interactions --- naturally appears in  higher-derivative theories with non-Abelian gauge structure and give origin at most to logarithmic divergences as far as the radiative corrections are concerned. On the other hand, trivial power-counting arguments immediately show that the new higher-dimensional operators do not break renormalizability at all  since the improved ultraviolet behavior of the bosonic propagator $P(k)$ in the deep Euclidean region scales as $P(k) \sim \frac{1}{k^4}$, instead of the usual $P(k) \sim \frac{1}{k^2}$ when $k^2 \rightarrow \infty$.  It is worth noticing that every field in the Standard Model has a LW partner with an associated LW mass. These masses are the only new parameters in the minimal LWSM. However, as pointed out by \'Alvarez, Schat, Da Rold and Szynnkman, the LWSM does not provide any information on the origin of the LW  masses which, in order to solve the hierarchy problem, should not be heavier than a few TeV. In fact,  an analysis of the    electroweak precision observables in the LWSM  shows that in order to reproduce the electroweak data (assuming that all the  LW masses are of the same order) the  LW scale should be of order 5TeV \cite{4}.

Although  only two years have elapsed since the emergence of the LWSM, there exists already a large body of literature related to this subject \cite{5,6,7,8,9,10,11,12,13}, which clearly shows the considerable interest this model has aroused.  Phenomenological studies on the LWSM can also be found in abundance \cite{4,14,15,16,17,18}.   

Since the  LWSM is an extension of the LW model,  a   better understanding of    some basic features of the   latter --- such as order of magnitude of the  LW heavy photon mass, deviation of the behavior of static electromagnetic fields and a possible pacific coexistence between  the heavy photon and magnetic charge --- would certainly help us to  improve our knowledge of the former. In a sense, this type of research is, {\it mutatis mutandis}, similar to that conducted in lower-dimensional theories  with the purpose of gaining insight into difficult conceptional issues, which are present and even more opaque  in the physical (3+1)-dimensional world. 

Our goal in this article is precisely to examine the aforementioned aspects of the  LW finite QED. To start off, we shall undertake a brief review of the alluded model in  Section 2. Three bounds on the LW heavy photon mass are then estimated in Section 3. The first  limit is found using the data from an old but very accurate experiment carried out by Plimpton and Lawton to test the Coulomb's law of force between charges \cite{19}. The second bound  is estimated by taking into account that the point particle limit of the self-force acting on an extended nonrelativistic charged particle, in the framework of the LW finite QED at short distances, is finite and well defined \cite{20,21}. Finally, the third limit is obtained by computing the  anomalous electron magnetic moment within the context of the LW model.  In Section 4 we calculate the nonrelativistic potential for the interaction of two  fermions in the framework of the LW finite QED  via two powerful but quite different methods. The first approach is based on the merging of quantum mechanics with the nonrelativistic limit of quantum field theory, while the second one adopts a gauge-invariant but path-dependent variables formalism \cite{22,23,24,25}. Since both calculations lead to the same result, we may conjecture that these radically different methods are  equivalent term by term. To conclude, we analyze in Section 5 the possibility of occurrence of monopoles in the context of the LW model.

In our conventions $\hbar =c =1$ and the signature of the metric is $(+1,\;-1,\; -1,\; -1)$. 
\section{An overview of the LW finite electrodynamics}
The Abelian LW model is defined by the following gauge-invariant Lagrangian

\begin{eqnarray}
{\cal{L}} =- \frac{1}{4}F_{\mu \nu}F^{\mu \nu} - \frac{1}{4m^2}F_{\mu \nu}\Box F^{\mu \nu},
\end{eqnarray}

\noindent where $F_{\mu \nu} (= \partial_\mu A_\nu - \partial_\nu A_\mu)$ is the field strength.

Let us then show that the above Lagrangian describes two independent (on-shell) spin-1 fields:  massless one and  massive one, with positive and negative norm, respectively. To do that it is appropriate to provide another formulation where an auxiliary field is introduced and the higher-derivative term is absent. The field theory with real vectorial fields $A_\mu$ and $Z_\mu$ with Lagrangian

\begin{eqnarray}
{\cal{L}} &=& \frac{1}{2}A_\mu \Box Z^\mu + \frac{1}{2}\partial_\mu A^\mu \partial_\nu Z^\nu - \frac{m^2}{8}A_\mu A^\mu \nonumber \\ &&+ \frac{m^2}{4}A_\mu Z^\mu - \frac{m^2}{8}Z_\mu Z^\mu,
\end{eqnarray}

\noindent is equivalent to the field theory with the Lagrangian in Eq. (1). In fact, varying $Z_\mu$ gives

\begin{eqnarray}  
Z_\mu = A_\mu + \frac{2}{m^2}\Box A_\mu - \frac{2}{m^2}\partial_\mu \partial_\nu A^\nu,
\end{eqnarray}

\noindent and the coupled second-order equations from (2) are fully equivalent to the fourth-order equations from (1). The system (2) now separates cleanly into the Lagrangians for two fields, when we make the change of variables

\begin{eqnarray}
A_\mu = B_\mu + C_\mu, \\ Z_\mu= B_\mu - C_\mu.
\end{eqnarray}

\noindent In terms of $B_\mu$, $C_\mu$, $B_{\mu \nu} \equiv \partial_\mu B_\nu - \partial_\nu B_\mu$ and $C_{\mu \nu } \equiv \partial_\mu C_\nu - \partial_\nu C_\mu$, the Lagrangian now becomes

\begin{eqnarray}
{\cal{L}} = -\frac{1}{4}B_{\mu \nu}B^{\mu \nu} + \frac{1}{4}C_{\mu \nu}C^{\mu \nu} - \frac{m^2}{2}C_\mu C^\mu,
\end{eqnarray}

\noindent which is nothing but the difference of the Maxwell Lagrangian for $B_\mu$ and the Proca Lagrangian for $C_\mu$.

The particle content of the theory can also be obtained directly from Eq. (1). To accomplish this goal we compute the residues at the simple poles of the saturated propagator (contraction of  the propagator with conserved currents). Adding to (1) the gauge-fixing term ${\cal{L}}_\lambda = -\frac{1}{2\lambda}(\partial_\mu A^\mu)^2$, where as usual $\lambda$ plays the role of the gauge-fixing parameter, and noting that due to the  structure of the theory and the choice of a linear gauge-fixing functional, no Faddeev-Popov ghosts are required in this case, we promptly get the propagator in momentum space, namely,

\begin{eqnarray} 
D_{\mu \nu}(k) &=& \frac{m^2}{k^2(k^2- m^2)}\left\{\eta_{\mu \nu} - \frac{k_\mu k_\nu }{k^2} \Bigg[1  \right.  \nonumber \\   &&+  \left. \lambda \left(\frac{k^2}{m^2} - 1  \right) \Bigg]\right\}.
\end{eqnarray}

\noindent Contracting (7) with conserved currents $J^\mu(k)$, yields 

\begin{eqnarray}
{\cal{M}} \equiv J^\mu D_{\mu \nu} J^\nu, \nonumber \; \;= -\frac{J^2}{k^2} + \frac{J^2}{k^2 - m^2},
\end{eqnarray}

\noindent which allows us to conclude, taking into account that $J^2 < 0$ \cite{26,27}, that the signs of the residues of ${\cal{M}}$ at the poles $k^2=0$ and $k^2 = m^2$ are, respectively, 

$$Res {\cal{M}}(k^2=0) >0, \;\;\; Res{\cal{M}}(k^2=m^2)<0,$$ 

\noindent which confirms our previous result.

It is worth noticing that the wrong sign of the residue of the heavy photon indicates an instability of the theory at the classical level. From the quantum point
of view it means that the theory is nonunitary. Luckily, these difficulties can be circumvented. Indeed, the classical instability can be removed by imposing a future boundary condition  in order to prevent exponential  growth of certain modes. However, this procedure leads to causality violation in the theory \cite{28}; fortunately, this acausality is suppressed below the scales associated with the LW particles. On the other hand, Lee and Wick argued that despite the presence of the aforementioned degrees of freedom associated with a non-positive definite norm on the Hilbert space,  the theory could nonetheless be unitary as long as the new LW particles obtain decay widths. There is no general proof of unitary at arbitrary loop order for the LW electrodynamics; nevertheless, there is no known example of unitarity violation. Accordingly, the LW electrodynamics is finite. Therefore, we need not be afraid of the massive spin-1 ghost.   

In summary, we may say that  the LW work consists essentially in  the introduction of  Pauli-Villars, wrong-sign propagator, fields as physical degrees of freedom which leads to amplitudes that are better behaved in the ultraviolet and render the logarithmically divergent QED finite.  

We remark that for the sake of convenience we shall work in the representation of the gauge field $A_\mu$ as given in Eq. (1), with the propagator   as in  Eq. (7).  

\section{Bounding the LW heavy photon mass}
We shall now estimate three bounds on the LW heavy photon mass. The first two limits will be found treating the LW model as a classical electromagnetic system,
 while the third one will be obtained considering the LW system as a quantum field model. One of the classical bounds is based on  the measurements obtained on a lab experiment carried out by Plimpton and Lawton  with the aim of testing the Coulomb inverse square law, the other relies upon the fact that under certain circumstances the  LW electrodynamics at short distances solves the  famous $\frac{4}{3}$ problem of the Maxwell's electromagnetism. The quantum bound, in turn, involves the computation of the anomalous electron magnetic moment in the framework of the LW finite electrodynamics.    

\subsection{The Plimpton-Lawton experiment}
In the presence of a source, ${\cal{L}}_\mathrm{source}= - A^\mu J_\mu$, the field equations concerning the LW model are

\begin{eqnarray}
(1+l^2\Box)\partial_\mu F^{\mu \nu}&=& j^\nu, \\ \partial_\mu \tilde{F}^{\mu \nu} &=&0,
\end{eqnarray}

\noindent where $\tilde{F}^{\mu \nu} = \frac{1}{2} \varepsilon^{\mu \nu \alpha \beta}F_{\alpha \beta} \;(\varepsilon^{0123}= +1)$, and $l\equiv m^{-1}$. In the limit of static charge Eq. (8) reduces to

\begin{eqnarray}
(1-l^2 \nabla^2)\nabla^2 \phi = - \rho,
\end{eqnarray}

\noindent with ${\bf E}= - \nabla \phi$. The parameter $l$ (which works as a cutoff) introduces a natural length scale for electrostatics.

Now, in the Plimpton-Lawton experiment \cite{19}, it was detected a voltage difference $\Delta V$ between a conducting sphere of a radius $R_2$ raised to a high  potential $V$ and a second concentric uncharged sphere of radius $R_1$ contained in the first one. From Eq. (10) one has

\begin{eqnarray}
\phi(r)= \frac{VR_2 \left[2r + l \left(e^{- \frac{R_2+r}{l}} - e^{- \frac{R_2-r}{l}}\right) \right]}{\left[2R_2 + l\left(e^{-\frac{2R_2}{l}}-1\right)\right]r}\nonumber
\end{eqnarray}

\noindent between the spheres, and hence

\begin{eqnarray}
&&\frac{\Delta V}{V} \equiv \frac{\phi(R_2) -\phi(R_1)}{\phi(R_2)}, \nonumber \\ &=&
1- \frac{ R_2 \left[ 2R_1 + l \left( e^{- \frac{R_2 + R_1}{l}} - e^{-\frac{R_2 - R_1}{l}} \right)\right]}{R_1\left[2R_2 + l\left( e^{-\frac{2R_2 }{l}} -1 \right) \right] }. \nonumber
\end{eqnarray}

Now, taking into account that both $\frac{R_1}{l}$ and $\frac{R_2}{l}$ are $\gg 1$, we promptly obtain

\begin{eqnarray}
\frac{\Delta V}{V} \approx - \frac{l}{2R_2}. 
\end{eqnarray}

On the other hand, in the original experiment conducted by Plimpton and Lawton, a harmonically alternating potential of $3000V$ was applied to the outer sphere. Tests were then made to detect the change in potential of the inner sphere relative the outer one. The results showed that no change in the thermal driven motion of the galvanometer could be detected during the course of the experiment, at a detector sensitivity  $1 \mu V$. The radii of the two spheres were $0.76m$ and $0.61m$, respectively. Substitution of the experiment parameters into Eq. (11) yields a limit on the cutoff  $l < 5.1\times 10^{-10} m$. 

Many improvements to the experiment of Plimpton and Lawton were reported in the '60s-'80s \cite{29,30}. In all of them, however, the  distance between measurement points were, roughly speaking, greater than that utilized in the early  experiment. Now, since the cutoff $l$ introduces a natural length scale for electrostatics, as we have already commented, obviously for a given precision of field measurement, the longer the distance between measurement points, the worst the cutoff limit obtained. Therefore, we come to the conclusion that  among the  experiments under discussion, the original experiment of Plimpton and Lawton is just the very one that provides the best upper bound on $l$ and, as a consequence, the best  lower bound on the LW heavy photon mass, i.e.,  $m \approx 0.38 eV$. 

\subsection{$\frac{4}{3}$ problem in classical electrodynamics}

The Newton equation of motion of an extended charge in the framework of the Abraham-Lorentz model takes the form

\begin{eqnarray}
\frac{4}{3} m_e \dot{{\bf v}} - \frac{2}{3}e^2 \ddot{{\bf v}} = {\bf F}_\mathrm{ext},
\end{eqnarray} 

\noindent where ${\bf v}$ and  $m_e$ are, respectively, the velocity and the electromagnetic mass of the particle and ${\bf F}_\mathrm{ext}$ is the external force applied to the electron. Here it is assumed that in the instantaneous rest frame of the particle, the charge distribution is rigid and spherically symmetric. Besides, it is also supposed the absence of mechanical mass. 

Of course, the electromagnetic mass enters with an incorrect coefficient in Eq.(12). For this reason the factor $\frac{4}{3}$ has been the source of considerable trouble. A possible way  out of this difficult, among others, is to evaluate the self-force acting on a point charged particle in the  framework of the LW finite electrodynamics at short distances. In this case the electromagnetic mass occurs in the equation of motion in a form consistent with especial relativity and, in addition, the exact equation of motion does not exhibit runaway solution or non-casual behavior, when the cutoff $l$ is larger than half of the classical radius of the electron \cite{20,21}.  

Taking the previous considerations into account, we promptly obtain $l< 1.4 \times 10^{-15} m$. Accordingly, a lower bound on the LW heavy photon mass is $m \approx 140 MeV$. 

\subsection{The  anomalous electron magnetic moment}
The preceding results suggest that a better bound on $l$ could be obtained in the quantum realm. Now, taking into account that QED predicts the anomalous magnetic moment of the electron correctly to ten decimal places, a quantum bound on $l$ can be found by computing the anomalous magnetic moment of the electron in the context of the LW electrodynamics and comparing afterward the result obtained with that of QED. To accomplish this goal, we recall that the anomalous magnetic moment of the electron stems from the vertex correction for the scattering of the electron by an external field, as it is shown in Fig. 1.

\begin{figure}[!hbp]
\begin{center}
\includegraphics[scale=0.58]{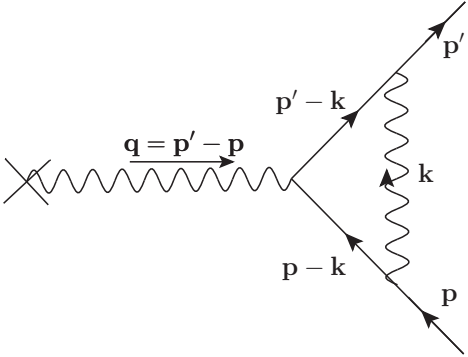}
\end{center}
\caption{\small Vertex correction for electron  scattering by  an external  field.\label{fig1}}
\end{figure}

For an electron scattered by an external static magnetic field and in limit ${\bf q} \rightarrow {\bf 0}$, the gyromagnetic ratio is given by\cite{31}

$$g = 2[1 + 2F_2(0)].$$

\noindent The form factor of the electron, $F_2 (0)$, corresponds to a shift in the $g-$factor, usually quoted in the form $F_2(0)\equiv \frac{g-2}{2}$, and yields the anomalous magnetic moment of the electron.  By employing (7) in the calculation of the diagram in Fig. (1), it can be shown that

\begin{eqnarray}
F_2 (0) &=& \frac{\alpha}{\pi}\int_0^\infty d\alpha_1 d\alpha_2 d \alpha_3 \delta (1 - \Sigma \alpha_i) \nonumber \\  &&\times \left[\frac{\alpha_1}{\alpha _2 + \alpha_3}   -   \frac{\alpha_1^2 (\alpha_2 + \alpha_3)}{(\alpha_2 + \alpha_3)^2 + \frac{\alpha_1}{\varepsilon}} \nonumber \right],
\end{eqnarray}

\noindent where $\varepsilon \equiv l^2 M^2$, $M$ being the electron mass. We call attention to the fact that the term $-\frac{k^\mu k^\nu}{k^4(l^2 k^2 -1)}[1 + \lambda(l^2 k^2-1)]$ that appears in Eq. (7) makes  no  contribution to the the form factor $F_2 (0)$ because the propagator always occurs coupled to conserved currents.

Integrating the above expression  first with respect to $\alpha_3$ and subsequently with respect to $\alpha_2$, gives

\begin{eqnarray}
F_2 (0) &=& \frac{\alpha}{\pi}\int_0^1 d\alpha_1 \int_0^{1- \alpha_1} d \alpha_2\left[\frac{\alpha_1}{1- \alpha_1} \right. \nonumber \\ &&- \left.  \frac{\alpha_1 (1 - \alpha_1)}{(1- \alpha_1)^2 + \frac{\alpha_1}{\varepsilon}} \right] \nonumber \\ &=& \frac{\alpha}{\pi}\int_0^1 d \alpha_1 \frac{\alpha_1^2}{\alpha_1 + \varepsilon (1 -\alpha_1)^2}. \nonumber 
\end{eqnarray}

Now,

\begin{eqnarray}
\int dx \frac{x^2}{\varepsilon x^2 + x(1 - 2 \varepsilon) + \varepsilon}=\frac{x}{\varepsilon} - \nonumber \\ \frac{1- 2\varepsilon}{2 \varepsilon}\ln \left|\varepsilon x^2 + x(1 -2\varepsilon) + \varepsilon \right|\nonumber \\ + \frac{1 + 2\varepsilon^2 - 4\varepsilon}{2\varepsilon^2 \sqrt{1 -4\varepsilon}} \nonumber  \ln \left|\frac{A-B}{A+B}\right|, \nonumber 
\end {eqnarray}

\noindent where $A \equiv 2\varepsilon x +1 - 2\varepsilon$, and $B \equiv   \sqrt{1 - 4\varepsilon}$; hence

\begin{eqnarray}
F_2(0)  &=&\frac{\alpha}{\pi} \left[ \frac{1}{\varepsilon} + \frac{1 - 2\varepsilon}{2\varepsilon^2}\ln \varepsilon +\frac{1 + 2\varepsilon^2 - 4\varepsilon}{2\varepsilon^2 \sqrt{1-4\varepsilon}} \right. \nonumber \\ &&\times  \left.\ln \frac{1 + \sqrt{1- 4\varepsilon}}{1 - \sqrt{1-4\varepsilon}} \right].
\end{eqnarray}

Recalling that $\varepsilon \ll 1$, which implies that

$$\frac{1 + 2\varepsilon^2 - 4\varepsilon}{2\varepsilon^2 \sqrt{1-4\varepsilon}} \approx \frac{1 - 2\varepsilon + 5\varepsilon^4}{2\varepsilon^2}, $$ 

\begin{eqnarray}
\ln \frac{1 + \sqrt{1- 4\varepsilon}}{1 - \sqrt{1-4\varepsilon}} &\approx& -\bigg[ 2\varepsilon + 3\varepsilon^2 + \frac{20\varepsilon^3}{3} + \frac{35\varepsilon^4}{2}  \nonumber \\  &&+  \ln \varepsilon \bigg],  
\end{eqnarray}

\noindent  we arrive at the conclusion that

\begin{eqnarray}
F_2(0) &\approx& \frac{\alpha}{2 \pi}\bigg[ 1 -\frac{2}{3}(lm)^2 - 2\left( \frac{25}{12} +  \ln (lm) \right)\nonumber \\ &&\times(lm)^4 +  {\cal O}\left((lm)^6\right) \bigg].
\end{eqnarray}

\noindent The first term of the above equation is equal to that calculated by Schwinger in 1948 \cite{32}. Since then  $F_2(0)$ has been calculated to order $\alpha^8$ for QED. The second term of Eq. (15) is the most important correction related to the parameter $l$ of the LW electrodynamics. 
Recent calculations  concerning $F_2(0)$ in the framework of QED give for the  electron \cite{33} 

$$F_2(0) = 1 \;159 \;652\; 182.79(7.71)\times 10^{-12},$$

\noindent where the uncertainty comes mostly from that of the best non-QED value of the  fine structure constant $\alpha$. The current experimental value for the anomalous magnetic moment is, in turn \cite{34},

$$F_2(0) = 1 \;159 \;652\; 181.1(0.7)\times 10^{-12}.$$

\noindent Comparison of the theoretical value predicted by QED  with the experimental one shows that these results agree in $1$  part in  $10^{10}$. As a consequence, 

$$\frac{2}{3} (lm)^2 < 10^{-10},$$

\noindent implying $l < 4.7 \times 10^{-18} m$. Consequently, a lower limit on the heavy photon Lee and Wick hypothesized the existence is $m \approx$ 42 GeV.

\subsection{Discussion}
The first classical bound on the LW heavy photon mass we have found is not a good limit since the distance between the measurement points is too large. It would thus be interesting and instructive to estimate what value the radius of the   outer sphere of a Plimpton-Lawton-like experiment should have in order to provide a bound on the LW heavy photon mass of the order, say  of the vectorial bosons masses $(\approx 100 GeV)$. A simple order-of-magnitude calculation shows that $R_2 \approx 5.9 \times 10^{-9} m$, as long as we do not change neither the detector sensitivity nor the potential of the external sphere related to the original  experiment. Such an experiment, obviously, does not make sense. Accordingly, we shall not consider the first classical limit in our discussions.  We shall also discard the second classical bound since it relies heavily on the electron classical radius, a   parameter that cannot be obtained experimentally. The remaining bound, i.e., the quantum one, is extremely reliable  since it was established using one of the great triumphs of QED, namely the astonishing agreement between theory and experiment to ten decimal places for the anomalous electron magnetic moment. Besides, this limit is based on truly (loop)  quantum effects. Therefore, from now on we shall assume in our considerations that  the mass of the LW heavy photon mass is equal to that estimated via  the  quantum bound ($\approx 42 GeV$).  We remark that this value is within the allowed range of $7-90 \;GeV$ for $m$ estimated by Lee and Wick \cite{2} by requiring consistence with known data; it is also of the same order as the value for $m$ considered by Resnick and Sundaresan  $(22.1GeV)$ in their calculation on the possible effects of the LW heavy photon in $e^- e^+$ elastic scattering \cite{35}. It is amazing that the bound on the mass of the LW heavy photon we have derived is of the order of the masses of the vectorial bosons found in nature. ($m_\mathrm{W} \approx 80GeV, \; m_\mathrm{Z} \approx 91GeV$ \cite{34}). It is also worth noticing that the masses of the LW heavy particles are in the GeV scale, while those related  to the LWSM are in the TeV scale; being, as a result, three orders of magnitude lower than those concerning the LWSM,  which is fairly reasonable. 

According to the quantum bound estimated on  the LW heavy particle,  the latter could, in principle, have been detected at the LEP. However, we have to admit that the scenario analyzed which yields this estimate at the GeV scale is such that the higher-derivative term is added up to the photon,  already separated from the Z$^0$ gauge boson after spontaneous symmetry breaking has occurred.  Nevertheless, the common origin of electromagnetic and weak interactions, as realized in the $SU(2)\times U(1)$ Electroweak Theory, tells us that the dimension-six term, once introduced into the Lagrangian prior to the spontaneous  $SU(2)\times U(1)$-symmetry breakdown, must be governed by a mass$^{-2}$-dimensional parameter which should better be  connected  to a parameter originated from a physics beyond the SM. It may be the outcome of a large extra dimension and this would more naturally accommodate the result of a heavy photon at the Tev scale, appropriate to account for BSM (Beyond Standard-Model) Physics. In this context, it would be advisable to  reassess the Z$^0$' physics in the framework of the LW Electroweak Theory.

\section{Coulomb's law modification in the framework of the LW finite electrodynamics}
We shall now study the lowest-order modification of the Coulomb potential due to  the higher-order terms coming from the LW finite QED,  through two complete different methods. The first approach is based on the marriage of quantum mechanics with the nonrelativistic limit of quantum field theory, whereas the second one makes use of a gauge-invariant but path-dependent variables formalism. A remarkable feature of the latter method is that it provides a physically-based alternative to the usual Wilson loop approach.

\subsection{Merging quantum mechanics with the nonrelativistic  limit of quantum field theory }
Consider the scattering of two fermions with identical masses in the framework of the LW QED; specifically, an electron ($e^-$) and a positron ($e^+$), as it is depicted in Fig. 2.  Now, according to the Born approximation in the Schr\"odinger theory the differential cross section for the scattering of two particles with the same mass, $M$, is $\left( \frac{d\sigma}{d \Omega}  \right)_\mathrm{C.M.}= \left|\frac{M}{4\pi} \int{e^{- {\bf Q}\cdot {\bf r}}U(r)d^3{\bf{r}}}\right|^2 $. (The potential energy $U(r)$ causes the momentum transfer ${\bf Q}= {\bf p}' -{\bf p}$, where ${\bf p}$ and ${\bf p}'$ are the initial and final momenta of  one of the particles.) On the other hand, in the nonrelativistic limit the differential cross section for the interaction of the abovementioned fermions   can be expressed as  $\left( \frac{d\sigma}{d \Omega}  \right)_\mathrm{C.M.}= \left|\frac{{\cal M}_\mathrm{N.R.}}{16\pi M} \right|^2 $, wherein ${\cal M}_\mathrm{N.R.}$ is the nonrelativistic limit of the Feynman amplitude for the process at hand. As a result, in terms of the the momentum exchanged ${\bf k}= - {\bf Q}$  the nonrelativistic potential assumes the form 
 
\begin{eqnarray}
U(r)= \frac{1}{4M^2} \frac{1}{(2\pi)^3} \int{d^3{\bf{k}} {\cal M}_\mathrm{N.R.} e^{-i{\bf k}\cdot{\bf r}}}.
\end{eqnarray}

\noindent This formula enables us to construct an effective three-dimensional potential (to be used in connection with the Shr\"odinger equation as well as in the classical mechanics) once we know the nonrelativistic limit of the covariant matrix element.

\begin{figure}
\begin{center}
\includegraphics[scale=0.4]{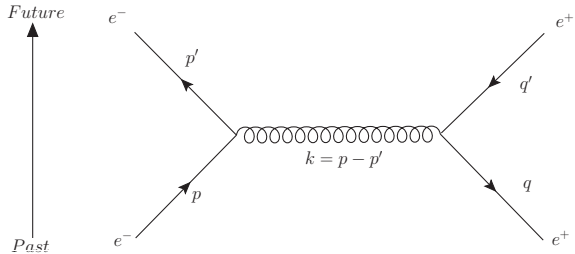}
\end{center}
\caption{\small Lowest order contribution to elastic $e^-e^+$ scattering in the context of the LW electrodynamics.\label{fig1}}
\end{figure}

 On the other hand, from the Lagrangian for the interaction of the mentioned particles, i.e, ${\cal L}_\mathrm{int}= e{\bar{\psi}} \gamma^\mu \psi A_\mu$, we promptly obtain   the Feynman rule for the elementary vertex (See Fig. 3). We remark that in our convention the  electron charge is equal to $-e$.  Accordingly, the invariant amplitude for the process shown in Fig. 2 is

\begin{eqnarray}
{\cal{M}} = -\frac{e^2 m^2}{k^2(k^2 - m^2)} {\bar{u}}(p') \gamma^\mu u(p) {\bar{v}}(q) \gamma_\mu v(q').
\end{eqnarray}

The leading term in the nonrelativistic approximation can be obtained by using the zero-momentum solutions for the spinors, namely,

\begin{eqnarray}
 {\bar{u}}_r (p')\gamma^\mu u_s(p) &\stackrel{\textrm{\tiny{N.R.}}}{\longrightarrow}&{\bar{u}}_r(0)\gamma^\mu u_s(0),\\
{\bar{v}}_r (p') \gamma^\mu v_s(p)&\stackrel{\textrm{\tiny{N.R.}}}{\longrightarrow}&{\bar{v}}_r(0)\gamma^\mu v_s(0),
\end{eqnarray}

\noindent with the normalization $ u_r^\dagger (p)u_s(p)=v_r^\dagger (p)v_s(p) = 2E_p \delta_{rs}$ \cite{36}. Consequently,

\begin{eqnarray}
{\bar{u}}_r(0)\gamma^0 u_s(0)&=& {\bar{v}}_r(0)\gamma^0 v_s(0)= 2M \delta_{rs}, \\ {\bar{u}}_r(0)\gamma^i u_s(0)&=& {\bar{v}}_r(0)\gamma^i v_s(0)= 0.
\end{eqnarray}

As a result,

\begin{eqnarray}
{\cal M}_\mathrm{N.R.} = - \frac{4e^2 m^2 M^2}{{\bf {k}}^2 ({\bf{k}}^2 + m^2)}.
\end{eqnarray}

Inserting (22) into (16)
and performing the resulting integral, we come to the conclusion that the effective potential we are searching for is

\begin{eqnarray}
U(r) = - \frac{e^2}{4\pi r}\left(1- e^{-mr}\right).
\end{eqnarray}

\begin{figure}[h]
\begin{center}
\includegraphics[scale=0.4]{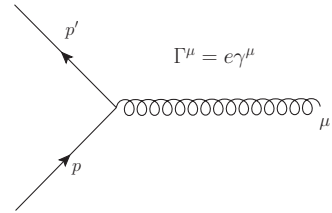}
\end{center}
\caption{\small 
Relevant vertex for the LW finite QED.\label{fig1}}
\end{figure}

\subsection{Computing the effective potential via a gauge-invariant but path-dependent variables formalism}

We shall now calculate the  nonrelativistic potential   using this time a gauge-invariant
but path-dependent variables formalism along the lines of Refs. \cite{22,23,24,25}, which, as we have already commented, is a physically-based
alternative to the usual Wilson loop approach. To this end,
we will compute the expectation value of the energy operator $H$ in the
physical state $|\Phi\rangle$ describing the sources, which we will denote by
${\langle H\rangle}_\Phi$.  In order to obtain the  Hamiltonian corresponding to the LW model, we must perform beforehand the
quantization of the theory. In this vein, we start off by computing
the  canonical momenta $\Pi ^\mu   =  - \frac{1}{{m ^2 }}\left( {m ^2
+ \Box } \right)F^{0\mu }$.  This yields the usual primary constraint  $\Pi^{0}=0$,
while the momenta are $\Pi ^i   = \frac{1}{{m ^2 }}\left( {m ^2  + \Box} \right)E^{i}$. Therefore, the canonical Hamiltonian takes the form
\begin{eqnarray}
H_C  &=& \int {d^3 x} \left\{ {\Pi ^i \partial _i A_0  - \frac{1}{2}\Pi ^i \frac{{m ^2 }}{{\left( {\Box  + m^2 } \right)}}\Pi _i } \right\} \nonumber\\
&+& \int {d^3 x} \frac{1}{4}F_{ij} \left( {\frac{{m ^2  + \Box }}{{m ^2 }}} \right)F^{ij}. \label{LW15}
\end{eqnarray}
Temporal conservation of the primary constraint, $\Pi_0$, leads to the
secondary constraint, $\Gamma_1 \left(x \right) \equiv \partial _i \Pi ^i=0$.
It is straightforward to check that there are no further constraints in the
theory. Consequently, the extended Hamiltonian that generates translations
in time then reads $H = H_C + \int {d^3 }x\left( {c_0 \left( x
\right)\Pi _0 \left( x \right) + c_1 \left( x\right)\Gamma _1 \left(
x \right)} \right)$. Here $c_0 \left( x\right)$ and $c_1 \left( x
\right)$ are arbitrary Lagrange multipliers. Moreover, it follows from
this Hamiltonian that $\dot{A}_0 \left( x \right)= \left[ {A_0
\left( x \right),H} \right] = c_0 \left( x \right)$, which is an arbitrary
function. Since $\Pi^0 = 0$ always, neither $ A^0 $ nor $ \Pi^0 $ are of
interest in describing the system and may be discarded from the theory.
As a result, the Hamiltonian becomes
\begin{eqnarray}
H &=& \int {d^3 x} \left\{ {c(x) \partial _i \Pi ^i - \frac{1}{2}\Pi ^i \frac{{m ^2 }}{{\left( {\Box  + m ^2 } \right)}}\Pi _i } \right\} \nonumber\\
&+& \int {d^3 x} \frac{1}{4}F_{ij} \left( {\frac{{m ^2  + \Box }}{{m ^2 }}} \right)F^{ij}, \label{LW20}
\end{eqnarray}
where $c(x) = c_1 (x) - A_0 (x)$.

The quantization of the theory requires the removal of non-physical
variables, which is accomplished by imposing a gauge condition such that the
full set of constraints becomes second class. A particularly
convenient choice is 
\begin{equation}
\Gamma _2 \left( x \right) \equiv \int\limits_{C_{\xi x} } {dz^\nu }
A_\nu \left( z \right) \equiv \int\limits_0^1 {d\lambda x^i } A_i
\left( {\lambda x} \right) = 0, \label{LW25}
\end{equation}
where  $\lambda$ $(0\leq \lambda\leq1)$ is the parameter describing
the spacelike straight path $ x^i = \xi ^i  + \lambda \left( {x -
\xi } \right)^i $, and $ \xi $ is a fixed point (reference point).
There is no essential loss of generality if we restrict our
considerations to $ \xi ^i=0 $. The choice (26) leads
to the Poincar\'e gauge \cite{22,37}. As a consequence,
the only nontrivial Dirac bracket for the canonical variables is
given by
\begin{eqnarray}
\left\{ {A_i \left( x \right),\Pi ^j \left( y \right)} \right\}^ *
&=&\delta{ _i^j} \delta ^{\left( 3 \right)} \left( {x - y} \right)
\nonumber\\
&-& \partial _i^x \int\limits_0^1 {d\lambda x^j } \delta ^{\left( 3
\right)} \left( {\lambda x - y} \right). \label{LW30}
\end{eqnarray}

We are now equipped to compute the interaction energy for the model
under consideration. As mentioned before, to do that we need to know  the expectation value of the energy operator
$H$ in the physical state $|\Phi\rangle$. Following Dirac \cite{38},  we write the physical
state $|\Phi\rangle$  as
\begin{eqnarray}
\left| \Phi  \right\rangle &\equiv& \left| {\overline \Psi  \left(
\bf y \right)\Psi \left( {\bf y}\prime \right)} \right\rangle \nonumber\\
&=&
\overline \psi \left( \bf y \right)\exp \left(
{iq\int\limits_{{\bf y}\prime}^{\bf y} {dz^i } A_i \left( z \right)}
\right)\psi \left({\bf y}\prime \right)\left| 0 \right\rangle,
\label{LW35}
\end{eqnarray}
where the line integral is along a spacelike path on a fixed time
slice, $q$ is the fermionic charge, and $\left| 0 \right\rangle$ is the physical vacuum state.
Note that the charged matter field together with the electromagnetic
cloud (dressing) which surrounds it, is given by
$\Psi \left( {\bf y} \right) = \exp \left( { - iq\int_{C_{{\bf \xi}
{\bf y}} } {dz^\mu A_\mu  (z)} } \right)\psi ({\bf y})$. Thanks to
our path choice, this physical fermion then becomes $\Psi \left(
{\bf y} \right) = \exp \left( { - iq\int_{\bf 0}^{\bf y} {dz^i  }
A_{i} (z)} \right)\psi ({\bf y})$. In other words, each of the
states ($\left| \Phi  \right\rangle$) represents a
fermion-antifermion pair surrounded by a cloud of gauge fields to
maintain gauge invariance.
Taking  the above Hamiltonian structure into account, we see that
\begin{eqnarray}
\Pi _i \left( x \right)\left| {\overline \Psi  \left( \bf y
\right)\Psi \left( {{\bf y}^ \prime  } \right)} \right\rangle  &=&
\overline \Psi  \left( \bf y \right)\Psi \left( {{\bf y}^ \prime }
\right)\Pi _i \left( x \right)\left| 0 \right\rangle \nonumber\\
&+&  q\int_ {\bf
y}^{{\bf y}^ \prime  } {dz_i \delta ^{\left( 3 \right)} \left( {\bf
z - \bf x} \right)} \left| \Phi \right\rangle. \nonumber\\
\label{LW40}
\end{eqnarray}
Now, since the fermions are assumed to be
infinitely massive (static) we can substitute $\Box$ by
$-\nabla^{2}$ in Eq. (\ref{LW20}). Therefore,  $\left\langle H \right\rangle _\Phi$ can be written  as
\begin{equation}
\left\langle H \right\rangle _\Phi   = \left\langle H \right\rangle _0
+ \left\langle H \right\rangle _\Phi ^{\left( 1 \right)}  +
\left\langle H \right\rangle _\Phi ^{\left( 2 \right)}, \label{LW45}
\end{equation}
where $\left\langle H \right\rangle _0  = \left\langle 0
\right|H\left| 0 \right\rangle$. The $\left\langle H \right\rangle _\Phi ^{\left( 1 \right)}-$ and $\left\langle H \right\rangle _\Phi ^{\left( 2 \right)}-$
terms are given, respectively, by
\begin{equation}
\left\langle H \right\rangle _\Phi ^{\left( 1 \right)}  =  - \frac{1}{2}
\left\langle \Phi  \right|\int {d^3 x} \Pi _i \Pi ^i \left| \Phi  \right\rangle,
\label{LW50a}
\end{equation}
and
\begin{equation}
\left\langle H \right\rangle _\Phi ^{\left( 2 \right)}  = \frac{{1}}{2}
\left\langle \Phi  \right|\int {d^3 x} \Pi _i \frac{\nabla ^2}{{\left( {\nabla ^2  - m^2 } \right)}}\Pi ^i \left| \Phi  \right\rangle.  \label{LW50b}
\end{equation}
Using Eq. (\ref{LW40}), the $\left\langle H
\right\rangle _\Phi ^ {\left( 1 \right)}$- and $\left\langle H \right\rangle _\Phi ^
{\left( 2 \right)}$- terms can be rewritten as
\begin{eqnarray}
\left\langle H \right\rangle _\Phi ^{\left( 1 \right)}  &=&  - \frac{q^2}{2}
\int {d^3 x} \int_{\bf y}^{{\bf y}^
\prime  } {dz_i^ \prime  } \delta ^{\left( 3 \right)} \left( {{\bf x} -
{\bf z}^ \prime  } \right) \nonumber\\
&\times& \int_{\bf y}^{{\bf y}^ \prime  } {dz^i }
\delta ^{\left( 3 \right)} \left( {{\bf x} - {\bf z}} \right),
\label{LW55a}
\end{eqnarray}
and
\begin{eqnarray}
\left\langle H \right\rangle _\Phi ^{\left( 2 \right)} &=&  \frac{{q^2
}}{2}\int {d^3 } x\int_{\bf y}^{{\bf y}^{\prime}} {dz_i^{\prime }}
\delta ^{\left( 3 \right)} \left( {{\bf x} - {\bf z}^{\prime}  }
\right) \nonumber\\
&\times& \frac{{\nabla ^2 }}{{\left( {\nabla ^2  - m^2 } \right)}}
\int_{\bf y}^{{\bf y}^ \prime  } {dz^i } \delta ^{(3)} \left( {{\bf x} - {\bf z}} \right).   \label{LW55b}
\end{eqnarray}
Following an earlier procedure \cite{23,24}, we see that the
potential for two opposite charges, located at ${\bf y}$ and ${\bf y}'$, takes the form

\begin{eqnarray}
U(r) =  - \frac{{e^2 }}{{4\pi r }}\left( 1 - e^{ - mr}  \right),
\end{eqnarray}

\noindent where $\left|{\bf y} - {\bf y}'\right|=r$.

It is worth noting  that there is an alternative but equivalent
way of obtaining the result \cite{25}, which highlights certain
distinctive features of our methodology. We start by considering
\cite{25}:
\begin{equation}
U \equiv q\left( {{\cal A}_0 \left( {\bf 0} \right) - {\cal A}_0
\left( {\bf y} \right)} \right), \label{LW65}
\end{equation}
where the physical scalar potential is given by
\begin{equation}
{\cal A}_0 \left( {x^0 ,{\bf x}} \right) = \int_0^1 {d\lambda } x^i
E_i \left( {\lambda {\bf x}} \right), \label{LW70}
\end{equation}
with $i=1,2,3$. This follows from the vector gauge-invariant field
expression \cite{37}
\begin{equation}
{\cal A}_\mu  \left( x \right) \equiv A_\mu  \left( x \right) +
\partial _\mu  \left( { - \int_\xi ^x {dz^\mu  } A_\mu  \left( z
\right)} \right), \label{LW75}
\end{equation}
where, as in Eq. (\ref{LW25}), the line integral is along a
spacelike path from the point $\xi$ to $x$, on a fixed slice time.
The gauge-invariant variables (\ref{LW75}) commute with the sole
first constraint (Gauss' law), confirming in this way that these fields are
physical variables \cite{38}. Note that Gauss' law for the
present theory reads $\partial _i \Pi ^i  = J^0$, where we have
included the external current $J^0$ to represent the presence of two
opposite charges. For $J^0 \left( {t,{\bf x}} \right) = q\delta
^{\left( 3 \right)} \left( {\bf x} \right)$ the electric field is
given by
\begin{equation}
E^i  = q\partial ^i \left( {G\left( {\bf x} \right) - G^ \prime 
\left( {\bf x} \right)} \right), \label{LW80}
\end{equation}
where $G\left( {\bf x} \right) = \frac{1}{{4\pi }}\frac{1}{{|{\bf x}|}}$ and
  $G'\left( {\bf x} \right) = \frac{e^{ - m  \left| {\bf x} \right|}}{4\pi \left| {\bf x} \right|}$ are the
Green functions in three space dimensions. Finally, replacing this
result in (\ref{LW70}) and using (\ref{LW65}), we reobtain (34), i.e.,
\begin{equation}
U(r) =  - \frac{{e^2 }}{{4\pi r}}\left( 1 - e^{ - mr}  \right)
. \label{LW85}
\end{equation}

\subsection{Discussion}
It  is remarkable that two quite different methods have led to the  same expression for the  effective tridimensional potential. This astonishing result seems to indicate that to lower orders the two  approaches  might be equivalent order by order. However, whether  this is a correct conjecture, it is still an open question.

We discuss now the properties of the nonrelativistic potential $U(r)$ we have found.
\begin{enumerate}
	\item Unlike the Coulomb potential which is singular at the origin, $U$ is finite there ($U(0)= -\frac{e^2}{4\pi l}$). The fact that the potential is finite for $r \rightarrow 0$, it is a clear evidence that the self-energy and the electromagnetic mass of a point-like particle are finite in the LW model. Indeed, in this model, as we have already discussed,  not only the electromagnetic mass and the  self-force are finite, it  also provides an adequate  scenario for solving  the $\frac{4}{3}$  problem. 
\item When $\frac{r}{l}\gg 1$, $U$ reduces to the Coulomb potential.
\item Only for small distances ($\frac{r}{l} \ll 1$), does $U$ differs significantly from the Coulomb potential (See Fig. 4).  A quick glance at Fig. 4 it is enough to convince us that this distance is $\approx 10^{-17} m$, one order of magnitude higher than the $l$-value.  

\end{enumerate}
  
\begin{figure}[h]
\begin{center}
\includegraphics[scale=0.42]{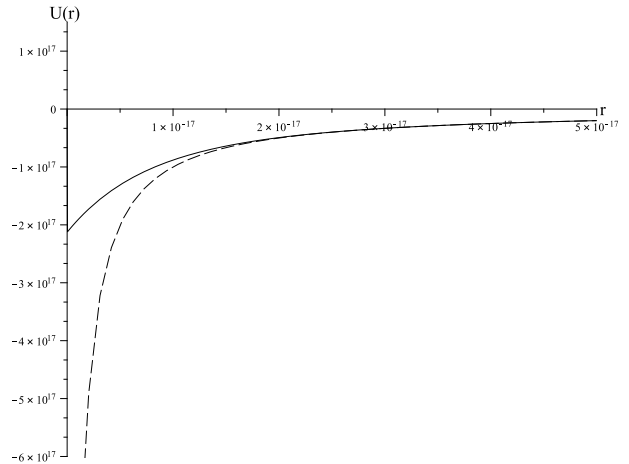}
\end{center}
\caption{\small 
The potential $U$ (in units of $\frac{e^2}{4 \pi}$), as a function of the distance $r$. The dashed line represents the Coulomb potential (in units of $\frac{e^2}{4 \pi}$).\label{fig1}}
\end{figure}

\section{Final Remarks}
An important question concerning the LW finite QED is whether or not the LW heavy photon and   magnetic charge can live in peace in its context. To answer this question we introduce a magnetic current $k^\mu = (\sigma, {\bf{k}})$ in the right-hand of Eq. (9). It is fairly straightforward to show that the resulting system of modified higher-order field equations, namely,

\begin{eqnarray}
(1 + l^2\Box)\partial_\mu F^{\mu \nu}&=&J^\nu, \\ \partial_\mu {\tilde{F}}^{\mu \nu}&=& k^\nu, 
\end{eqnarray}  

\noindent describes the existence of a magnetic charge. In fact, assuming the absence of electric fields, charges, and currents as well as the absence of magnetic current, we are left essentially with two equations for the magnetic field which have the familiar Dirac monopole solution ${\bf{B}} = \frac{g}{4\pi r^3}{\bf{\hat{r}}}$, where $g$ is  the magnetic charge. Using the usual methods, the Dirac quantization condition $\frac{qg}{4\pi}= \frac{n}{2}$, where $q$ is the electric charge, and $n$ is an integer, can be promptly recovered.

We have thus succeeded in finding a consistent system of Maxwell + vectorial boson mass  + magnetic charge equations.  We remark that the Dirac monopole and the massive vectorial boson cannot coexist in the context of Proca's massive electrodynamics \cite{39} because the latter, unlike the LW QED, is not gauge invariant. The very existence of the Dirac monopole is undoubtedly linked to the existence of the gauge invariance of the corresponding theory.

Interestingly enough the system formed by  Eqs. (41) and (42) is not symmetric under the duality transformation $F^{\mu \nu} \rightarrow \tilde{F}^{\mu \nu}, \; \tilde{F}^{\mu \nu} \rightarrow -F^{\mu \nu}, $ augmented by $j^\mu \rightarrow k^\mu, \; k^\mu \rightarrow
-j^\mu$. This fact raises an interesting question: Would it be possible to accommodate simultaneously magnetic charge and duality transformations in the framework  of a higher-order electromagnetic model?  A good attempt in this direction might be, for instance,  the model defined by the field equations

\begin{eqnarray}
(1 + l^2 \Box) \partial_\mu F^{\mu \nu} = J^\nu, \\(1 + l^2 \Box) \partial _\mu {\tilde F}^{\mu \nu} = k^\nu,
\end{eqnarray}

\noindent  since it is symmetric under duality transformations. It is worth noticing that $(1 + l^2 \Box) \partial _\mu {\tilde F}^{\mu \nu}$ is identically zero in the absence of the magnetic current.  Let us see then whether this model  admits a      monopole-like solution. For a    magnetostatic charge of strength $g$ fixed at the origin the solution of the  preceding equations is  ${\bf B}=  \frac{g}{4\pi}\left[ \frac{1 - e^{-r/l }}{r^2} - \frac{e^{-r/l}}{lr }\right] {\bf \hat{r}}$,  which for large distances reduces to  the Dirac result, as it should be.  Our point, nonetheless, is to ascertain whether or not this solution describes a  magnetic monopole at short distances. To see this we calculate the flux of the radial magnetic field  through a spherical surface $S$ of radius $r$ with the static monopole of strength $g$ at its center. Performing the computation we promptly find $\int_S {{\bf B} \cdot dS} = g\left[ 1 - \left( 1 + \frac{r}{l}\right) e^{-\frac{r}{l}}\right]$,  which implies that for $r/l \ll 1 $, $\int_S {\bf B \cdot dS} \approx 0$. Now, taking into account that if ${\bf B} =  \nabla \times {\bf A}$, $\int_S {\bf B \cdot dS}$ vanishes identically, we come to the conclusion that ${\bf A}$ can exist everywhere in the region under consideration. Therefore,  this  is a  monopole-like solution; nonetheless, the corresponding   magnetic charge, does not obey the Dirac quantization condition.  Indeed,  for $r/l \ll 1, \; {\bf B} \approx \frac{g}{4 \pi l^2} \frac{1}{r} {\bf r}$, implying that the magnetic field falls off like $1/r$ instead of $1/r^3$. This bizarre behavior of the magnetic field  certainly precludes us from recovering the Dirac  quantization condition.  One heuristic way of seeing that is to consider the motion of a particle of mass $m$ and charge $q$ in the field of the magnetic monopole. From the equation of motion of the particle,  $m{\bf \ddot{r}}= q{\bf\dot{r}}\times {\bf B}$, we get  the  ratio of change of its angular momentum,  $\frac{d}{dt}({\bf r} \times m {\bf \dot{r}})= \frac{qgr^2}{4 \pi l^2}  \frac{d}{dt} ({\bf \hat{r}})$, a result that prevents us from defining a conserved total angular momentum as in  the case of the Dirac monopole. Now, if the distances are neither too large  nor much small,  the potential vector cannot exist everywhere in the domain bounded by $S$ because ${\tilde F}^{\mu \nu} $ satisfies Eq. (44) rather than Eq. (42). Unlucky we cannot overcame this difficult  by introducing the concept of a string as Dirac did since in this case $\nabla \cdot {\bf B}$ ($ = \frac{g}{4 \pi} \frac{e^{- r/l}}{rl^2}$) does not vanishes anywhere in the aforementioned domain.
 
The preceding analysis leads us to conjecture that Dirac-like monopoles and duality transformations cannot be accommodated in the context of one and same higher-order electromagnetic model. 

To conclude, besides the reconsideration of the Z$^0$'-physics in the framework of the LW model, it would also be reasonable to reassess the issue of massive graviton production at LHC in the context of a LW version of gravity. Possible TeV-gravity signals at LHC, associated to the effects of large extra dimensions or a dynamical torsion in a BSM scenario, could be reproduced in a LW counterpart of gravity. This would allow to directly relate the mass$^{-2}$-parameter of the typical LW term to some large extra dimension and, in this context, it could perhaps become clearer the appearance of the GeV and the TeV mass scales for two generations of LW-type  heavy gravitons as well. An investigation of this particular problem is being pursued and we shall be reporting on its results elsewhere \cite {40}. 

Last but not least, we would like to draw the reader's attention to two recent articles by Fornal, Grinstein, and Wise (FGW) \cite{41}, and Cai, Qiu, Brandenberger, and Zhang \cite{42}, in which novel and important aspects of the LW model, are explored. In the first work \cite{41}, the high temperature behavior of the LW theories, including the LWSM, are analyzed. By examining the behavior of the LW resonances, FWG found that the contribution of the latter to the energy density and pressure is negative for high temperatures. In the second work, on the other hand, the cosmology  of a LW type scalar field field theory is considered and a nonsingular bouncing solution is obtained,   implying that the LW model provides a possible way out of the cosmological singularity problem. Besides, a scale invariant spectrum which can explain the recent CMB observations, is also found. These articles attest to the richness of the   LW models. Actually, the LW systems are the source of   a wealth of exciting new results  which deserve to be much better known.

\begin{acknowledgments}
A. Accioly, J. Helay\"el-Neto,  E. Scatena, and R. Turcati  are very grateful to FAPERJ, CNPq, and  CAPES (Brazilian agencies) for financial support. P. Gaete was supported in part by Fondecyt (Chile) grant 1080260.
\end{acknowledgments}


\begin{thebibliography}{99}
\bibitem{1}{T. Lee and G. Wick, Nucl. Phys. {\bf B9}, 209 (1969)}.
\bibitem{2}{T. Lee and G. Wick, Phys. Rev. D {\bf 2}, 1033 (1970)}.
\bibitem{3}{B. Grinstein, D. O'Connell, and M. Wise,  Phys. Rev. D {\bf 77}, 025012 (2008)}.
\bibitem{4}{E. \'Alvarez, C. Schat, L. Da Rold, and  A. Szynkman, J. Higher Energy Phys.   04 (2008) 026}.
\bibitem{5}{ C. Carone,  Phys. Lett. B {\bf 677}, 306 (2009)}.
\bibitem{6}{C. Carone and R. Lebed, J. Higher Energy Phys. 01 (2009) 043; Phys. Lett. B {\bf 668},  221 (2008)}.
\bibitem{7}{E. Gabrielli, Phys. Rev. D {\bf 77}, 055020 (2008)}.
\bibitem{8}{F. Wu and M. Zhong, Phys. Lett. B  {\bf 659}, 694 (2008); Phys. Rev. D {\bf 78}, 085010 (2008)}.
\bibitem{9} {A. Rodigast and  T. Schuster, Phys. Rev. D {\bf 79}, 125017 (2009)}.
\bibitem{10}{A. van Tonder, arXiv:0810.1928v1 [hep-th] 10 Oct 2008}.
\bibitem{11}{A. Shalaby, arXiv:0812.3419v2 [hep-th] 20 Jan 2009}.
\bibitem{12}{B. Grinstein, D. O'Connell, and M. Wise,  Phys.  Rev. D  {\bf 77}, 065010 (2008)}.
\bibitem{13}{I. Cho and O-Kab Kwon, arXiv: 1003.2716v1 [hep-th] 13 Mar 2010}.
\bibitem{14}{T. Underwood and R. Zwicky, Phys. Rev. D {\bf 79}, 035016 (2009)}.
\bibitem{15}{T. Rizzo, arXiv:0712.1791v2 [hep-th] 8 Jan 2008}.
\bibitem{16}{C. Carone and R. Primulando, Phys. Rev. D {\bf 80}, 055020 (2009)}.
\bibitem{17}{J. Espinosa, B. Grinstein, D. O'Connell, and M. Wise, arXiv:0705.1188v1 [hep-th] 9 May 2007}
\bibitem{18}{T. Rizzo,  arXiv:0704.3458v3 [hep-th] 11 Jun 2007}.
\bibitem{19}{S. Plimpton and W. Lawton, Phys. Rev.  {\bf 50}, 1066 (1936)}.
\bibitem{20}{J. Frenkel,  Phys. Rev. E  {\bf 54}, 5859 (1996)}.
\bibitem{21}{J. Frenkel and R. Santos, Int. J. Mod. Phys. B {\bf 13}, 315 (1999)}.
\bibitem{22} {P. Gaete and I. Schmidt, Phys. Rev. D {\bf61}, 125002 (2000).}
\bibitem{23}{P. Gaete and E. Spallucci, Phys. Rev. D {\bf 77}, 027702 (2008); J. Phys. A: Math. Theor. {\bf41}, 185401 (2008)}.
\bibitem{24}{P. Gaete and E. Spallucci, Phys. Lett. B {\bf675}, 145 (2009).}
\bibitem{25}{P. Gaete, Phys. Rev. D {\bf59}, 127702 (1999).}
\bibitem{26}{A. Accioly, Phys. Rev. D {\bf 67}, 127502 (2003); Nucl. Phys. B (Proc. Suppl.) {\bf 127}, 100 (2004).}
\bibitem{27}{A. Accioly and M. Dias, Int. J. Theor. Phys. {\bf 44}, 1123 (2005)}. 
\bibitem{28} {S. Coleman, in {\it Proceedings of Erice 1969, Ettore Majorana School on Subnuclear Phenomena} (Academic Press, New York, 1970), pp. 282-327.} 
\bibitem{29} {A. Goldhaber and M. Nieto, Rev. Mod. Phys.  {\bf 43}, 277 (1971).}
\bibitem{30} {Liang-Cheng Tu, J. Luo, and G. Gilles, Rep. Prog. Phys. {\bf 68}, 77 (2005).}
\bibitem{31}{P. Frampton, {\it Gauge Field Theories} (Benjamin/Cummings, California, 1987)}.
\bibitem{32}{J. Schwinger, Phys. Rev.  {\bf 73}, 416 (1948).}
\bibitem{33}{T. Aoyama, M. Hayakawa, T. Kinoshita, and M. Nio,  Phys. Rev. D {\bf 77}, 053012 (2008).}
\bibitem{34} {C. Amsler {\it et al.} (Particle Data Group), Phys. Lett. B {\bf667}, 1 (2008).}
\bibitem{35} {L. Resnick and M. Sundaresan, Phys. Rev. D {\bf5}, 264 (1972); {\bf 6}, 2721 (1972).}
\bibitem{36} {For an interesting discussion about representation-independent manipulations with Dirac matrices and spinors, see P. Pal, arXiv:physics/0703214v2 [physics.ed-ph] 10 Mar 2009.}
\bibitem{37} {P. Gaete, Z. Phys. C {\bf76}, 355 (1997).}
\bibitem{38} {P. Dirac, Can. J. Phys. {\bf33}, 650 (1955).}
\bibitem{39} {A. Ignatiev and G. Joshi, Phys. Rev. D {\bf53}, 984 (1996).}
\bibitem{40} {A. Accioly, P. Gaete, J. Helay\"{e}l-Neto, E. Scatena, and R. Turcati, work in progress.}
\bibitem{41}{B. Fornal, B. Grinstein, and M. Wise, Phys. Lett. B {\bf 674}, 330 (2009).}
\bibitem{42}{Yi-Fu Cai, T. Qiu, R. Brandenberger, and X. Zhang,  Phys. Rev. D {\bf 80}, 023511 (2009).}







\end{thebibliography}
\end{document}